# THE DECAYS $\tau \to \nu_\tau (K^*(892), K^*(1410), K_1(1270), K_1(1650), a_1(1260), a_1(1640))$ IN THE EXTENDED NAMBU – JONA-LASINIO MODEL

*M.K.Volkov*[1], *K.Nurlan*[1,2]

[1] Bogoliubov Laboratory of Theoretical Physics, JINR, Dubna 141980, Russia.
[2] Dubna State University, Dubna 141982, Russia
E-mail: volkov@theor.jinr.ru
nurlan.qanat@mail.ru

The $\tau \to \nu_\tau (K^*(892), K^*(1410), K_1(1270), K_1(1650), a_1(1260), a_1(1640))$ decays in the extended NJL model are calculated. The results obtained for $\tau \to \nu_\tau (K^*(892), K^*(1410), K_1(1270))$ decays are in a satisfactory agreement with the experimental data. There are no experimental data for other types of decay. Therefore, the obtained results should be considered as predictions for further experiments.

## Introduction

In the tau lepton decays, intermediate channels with vector and axial-vector mesons in the ground and first radial-exited states play a very important role. As a rule, for a description of those processes various phenomenological models, which describe the meson interaction at low energies, are used. One of the most effective models of this type is the extended NJL model. This model contains a minimal number of arbitrary parameters in comparison with other phenomenological models [1-5].

In the framework of this model the following processes were described: $\tau \to \nu_\tau \pi^- \pi^0$ [6], $\tau \to \nu_\tau \pi^- \eta(\eta')$ [7], $\tau \to \nu_\tau \pi^- \omega$ [8], $\tau \to \nu_\tau 2\pi \eta(550)(\eta'(950))$ [9], $\tau \to \nu_\tau K^- \pi^0$ [10], $\tau \to \nu_\tau K(\eta,\eta')$ [11], $\tau \to \nu_\tau K^0 K^-$ [12] where the channel with intermediate vector mesons $\rho$, $\rho'$, $K^*$, $K^{*'}$ play an important role. On the other hand, in the $\tau \to \nu_\tau 3\pi$ [13], $\tau \to \nu_\tau \pi^- \gamma$ [14], $\tau \to \nu_\tau f_1 \pi$ [15] decays the channel with intermediate axial-vector mesons plays the dominant role (see [16]). As intermediate vector and axial-vector channels play an important role in the calculation of the aforementioned processes, no doubt, the creation mechanism of vector or axial-vector mesons by means of tau lepton current is very interesting. This paper is devoted to the solution of this problem.

### 1. The Lagrangian of the extended NJL model for the vector $K^*(892)$, $K^*(1410)$ and axial-vector $K_1(1270)$, $K_1(1650)$, $a_1(1260)$, $a_1(1640)$ mesons.

In the extended NJL model, the quark-meson interaction Lagrangian for the vector $K^*(892), K^*(1410)$ and axial-vector $K_1(1270)$, $K_1(1650)$, $a_1(1260)$, $a_1(1640)$ mesons takes the form [10,17]:

$$\Delta L_{int}(q,\bar{q},K^*,K^{*'},a_1,a_1',K_1,K_1') = \bar{q}\frac{1}{\sqrt{2}}\gamma_\mu \{ \lambda_{a_1}\gamma_5(a_{a_1}a_{1\mu}^- + b_{a_1}a_{1\mu}^{'-}) +$$

$$+\lambda_K \left[ a_{K^*}(K_\mu^{*-} + K_{1\mu}^-) + b_{K^*}\gamma_5(K_\mu^{*'-} + K_{1\mu}^{'-}) \right] \} q \qquad (1)$$

where $\bar{q}, q$ - are the quark fields with masses $m_u = m_d = 280 MeV$, $m_S = 420 MeV$; $K^* = K^*(892)$, $K^{*'} = K^*(1410)$ are vector and $K_1 = K_1(1270)$, $K_1' = K_1(1650)$, $a_1 = a_1(1270)$, $a_1' = a_1(1650)$ are axial-vector mesons,

$$\lambda_K = \begin{pmatrix} 0 & 0 & 1 \\ 0 & 0 & 0 \\ 0 & 0 & 0 \end{pmatrix}, \quad \lambda_{a_1} = \begin{pmatrix} 0 & 0 & 0 \\ 1 & 0 & 0 \\ 0 & 0 & 0 \end{pmatrix} \qquad (2)$$

$$a_{K^*} = \frac{1}{\sin(2\beta_a^0)}\left[ g_{K^*}\sin(\beta_a + \beta_a^0) + g_{K^*}' f_a(\vec{k}^2)\sin(\beta_a - \beta_a^0) \right]$$

$$b_{K^*} = \frac{-1}{\sin(2\beta_a^0)}\left[g_{K^*}\cos(\beta_a+\beta_a^0)+g_{K^*}{}'f_a(\vec{k}^2)\cos(\beta_a-\beta_a^0)\right]$$

$$a_{a_1} = \frac{1}{\sin(2\alpha_a^0)}\left[g_\rho\sin(\alpha_a+\alpha_a^0)+g_\rho{}'f_a(\vec{k}^2)\sin(\alpha_a-\alpha_a^0)\right]$$

$$b_{a_1} = \frac{-1}{\sin(2\alpha_a^0)}\left[g_\rho\cos(\alpha_a+\alpha_a^0)+g_\rho{}'f_a(\vec{k}^2)\cos(\alpha_a-\alpha_a^0)\right] \quad (3)$$

$\beta_a = 84.74^0$, $\beta_a^0 = 59.56^0$ and $\alpha_a = 79.81$, $\alpha_a^0 = 61.44$ - are the mixing angles, $f_a(k^2) = 1+d_a k^2$ is the form factor for the description of the first radially excited states and $d_a$ - is the slope parameter ($d_{ud} = -1.784 GeV^{-2}$, $d_{us} = -1.761 GeV^{-2}$).

The coupling constants are:

$$g_{K^*} = \left(\frac{2}{3}I_2(m_u,m_s)\right)^{-1/2}, \quad g_{K^*}{}' = \left(\frac{2}{3}I_2^{f_{us}^2}(m_u,m_s)\right)^{-1/2}$$

$$g_\rho = \left(\frac{2}{3}I_2(m_u,m_u)\right)^{-1/2}, \quad g_\rho{}' = \left(\frac{2}{3}I_2^{f_{us}^2}(m_u,m_u)\right)^{-1/2} \quad (4)$$

The integral $I_2$ has the following form:

$$I_2^{f^n}(m_1,m_2) = -i\frac{N_c}{(2\pi)^4}\int\frac{f^n(\vec{k}^2)}{[m_1^2-k^2][m_2^2-k^2]}\theta(\Lambda_3^2-\vec{k}^2)d^4k \quad (5)$$

where $\Lambda_3 = 1.03\,GeV$ is the cut-of parameter

## 2. The amplitudes of the processes $\tau \to \nu_\tau(K^*, K^{*'}, K_1, K_1', a_1, a_1')$

The diagrams of the processes $\tau \to \nu_\tau(K^*, K^{*'}, K_1, K_1', a_1, a_1')$ are shown in Figs 1,2.

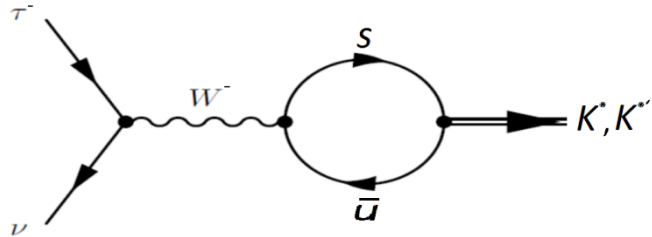

Figure1.The diagram describing decays $\tau \to \nu_\tau(K^*, K^{*'})$

($L_V^\nu = \bar{u}_\nu\gamma_\nu u_\tau$ - lepton current)

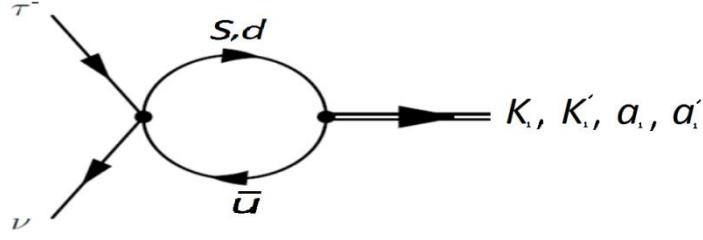

Figure2. The diagram describing decays $\tau \to \nu_\tau (K_1, K')$
$\tau \to \nu_\tau (a_1, a'_1)$, ($L^\nu_A = \bar{u}_\nu \gamma_5 \gamma_\nu u_\tau$ - lepton current)

The amplitudes of these processes take the form:

$$A_{\tau \to \nu_\tau K^*(K^{*'})} = \frac{G_F}{\sqrt{2}} L^\nu_V |V_{us}| \frac{1}{g_{K^*}} \left( p^2 g_{\nu\mu} - p_\nu p_\mu - \frac{3}{2}(m_s - m_u)^2 g_{\nu\mu} \right) e^K_\mu C_{K^*}(C_{K^{*'}}) =$$

$$= \frac{G_F}{\sqrt{2}} L^\nu_V |V_{us}| \frac{1}{g_{K^*}} \left( m^2_{K^*} - \frac{3}{2}(m_s - m_u)^2 \right) e^K_\nu C_{K^*}(C_{K^{*'}})$$

$$A_{\tau \to \nu_\tau K_1(K'_1)} = \frac{G_F}{\sqrt{2}} L^\nu_A |V_{us}| \frac{1}{g_{K^*}} \left( m^2_{K_1} - \frac{3}{2}(m_s + m_u)^2 \right) e^{K'}_\nu C_{K^*}(C_{K^{*'}})$$

$$A_{\tau \to \nu_\tau a_1(a'_1)} = \frac{G_F}{\sqrt{2}} L^\nu_A |V_{ud}| \frac{1}{g_\rho} \left( m^2_{a_1} - 6m_u^2 \right) e^a_\nu C_a(C_{a'}) \qquad (6)$$

where

$$C_{K^*} = \frac{\sin(\beta_a + \beta^0_a)}{\sin(2\beta^0_a)} + R_V \frac{\sin(\beta_a - \beta^0_a)}{\sin(2\beta^0_a)}, \quad C_{K^{*'}} = \frac{\cos(\beta_a + \beta^0_a)}{\cos(2\beta^0_a)} + R_V \frac{\cos(\beta_a - \beta^0_a)}{\cos(2\beta^0_a)}$$

$$C_a = \frac{\sin(\alpha_a + \alpha^0_a)}{\sin(2\alpha^0_a)} + R_B \frac{\sin(\alpha_a - \alpha^0_a)}{\sin(2\alpha^0_a)}, \quad C_{a'} = \frac{\cos(\alpha_a + \alpha^0_a)}{\cos(2\alpha^0_a)} + R_B \frac{\cos(\alpha_a - \alpha^0_a)}{\cos(2\alpha^0_a)} \qquad (7)$$

where

$$R_V = \frac{I^{f_{us}}_2(m_u, m_s)}{\sqrt{I_2(m_u, m_s) I^{f_{us} f_{us}}_2(m_u, m_s)}}, \quad R_B = \frac{I^{f_{ud}}_2(m_u, m_d)}{\sqrt{I_2(m_u, m_d) I^{f_{ud} f_{ud}}_2(m_u, m_d)}}. \qquad (8)$$

$G_F = 1.16637 \cdot 10^{-11} MeV^{-2}$ is the Fermi constant, $V_{us}, V_{ud}$ are elements of the Cabbibo-Kobayashi-Maskawa matrix

## 3. Numerical estimations

The square of the amplitude of the decay $\tau \to \nu_\tau K^*(892)$ takes the form:

$$\left|A_{\tau \to \nu_\tau K^*}\right|^2 = 2G_F^2 |V_{us}|^2 \frac{1}{g_{K^*}^2}\left(m_{K^*}^2 - \frac{3}{2}(m_s - m_u)^2\right)^2 \left(\frac{m_\tau^2 - m_{K^*}^2}{2}\right)\left(2 + \frac{m_\tau^2}{m_{K^*}^2}\right) C_{K^*}^2 \quad (9)$$

The squares of the amplitudes of other decays have an analogous form.

The decay width for the process $\tau \to \nu_\tau K^*(892)$ takes the form:

$$\Gamma(\tau \to \nu_\tau K^*) = \frac{\left|A_{\tau \to \nu_\tau K^*}\right|^2}{2 \cdot 2m_\tau} \int \frac{d^3 p_\nu}{2E_\nu (2\pi)^3} \frac{d^3 p_k}{2E_k (2\pi)^3} (2\pi)^4 \delta^4(p_\nu + p_k - p_\tau) \quad (10)$$

where the phase volume:

$$\int \frac{d^3 p_\nu}{2E_\nu (2\pi)^3} \frac{d^3 p_k}{2E_k (2\pi)^3} (2\pi)^4 \delta^4(p_\nu + p_k - p_\tau) = \frac{m_\tau^2 - m_{K^*}^2}{8\pi \cdot m_\tau^2} \quad (11)$$

The theoretical values and experimental data obtained for tau decay are given in the table.

| Decays | Theor NJL [MeV] | Epx PDG [MeV][19] |
|---|---|---|
| $\tau \to \nu_\tau K^*(892)$ | $2.60 \cdot 10^{-11}$ | $(2.72 \pm 0.15) \cdot 10^{-11}$ |
| $\tau \to \nu_\tau K^*(1410)$ | $5.15 \cdot 10^{-12}$ | $3,4\left(^{+3,178}_{-2,27}\right) \cdot 10^{-12}$ |
| $\tau \to \nu_\tau K_1(1270)$ | $0.907 \cdot 10^{-11}$ | $(1.06 \pm 0.244) \cdot 10^{-11}$ |
| $\tau \to \nu_\tau K_1(1650)$ | $6.78 \cdot 10^{-13}$ | — |
| $\tau \to \nu_\tau a_1(1260)$ | $3.20 \cdot 10^{-10}$ | — |
| $\tau \to \nu_\tau a_1(1640)$ | $1.437 \cdot 10^{-11}$ | — |

In this table, there are no theoretical and experimental data for the decay widths $\rho(770), \rho(1450)$, because this information is contained in the [18]. The decays $\tau \to \nu_\tau (K^*(892), K^*(1410))$ were also considered in that paper. However, large errors were made there. Therefore, in this paper, we again take into consideration these decays in the framework extended NJL model.

## 4. Conclusion

The experimental data of the considered here processes for tau lepton decays into the ground state and first-radially exited state are available only for the decays $\tau \to \nu_\tau (K^*(892), K^*(1410))$. Hence, we analyse the obtained theoretical results only for the $\tau \to \nu_\tau (K^*(892), K^*(1410))$ decays. The experiment shows that the $\tau \to \nu_\tau K^*(1410)$ decay width is 8 time less than the $\tau \to \nu_\tau K^*(892)$ decay width. In our model, this ratio is given by the dependence of the amplitude on the mixing angles of the ground and excited states (4 times). Further agreement with the experimental data is achieved due to the structure of the amplitude and describe in the phase volume.

In describing the transition of the lepton current into the vector meson many author use the model of vector dominance [20-24]. It is interesting to note that in our case the use of the model of vector dominance leads to the same results as it was obtained by means of MVD. At the same time, in more complicated decay processes of tau leptons, when vector mesons are intermediate states and decay into more complex final products, it is appears that multiplication of the $p_\nu p_\mu$ term by the top tends to be zero. This takes place for the processes $\tau \to \nu_\tau (\pi^- \pi^0, K^- K^0, \pi^- \omega)$. For decays such as $\tau \to \nu_\tau (\pi(\pi'(1300)), \pi\eta(\eta'), K\pi, K\eta(\eta'))$ we cannot neglect the $p_\nu p_\mu$ term. Therefore, our gradient invariant form of the amplitude for describing the transition of the lepton current into vector (axial-vector) meson, in most cases, coincides with MVD.

## Acknowledgements

We are grateful to A.B. Arbuzov and A.A. Pivovarov for useful discussions.